М.Л. Калужский

**Маркетинговые особенности дропшиппинга в системе электронной коммерции**

*Аннотация*: Дропшиппинг сегодня стремительно завоёвывает интернет и превращается в один из основных инструментов маркетинга в электронной коммерции. В статье раскрываются маркетинговые особенности, механизмы и значение дропшиппинга в условиях сетевой экономики XXI века. Автор проводит сравнительный анализ институционального развития дропшиппинга в США, Китае и России.

*Ключевые слова*: электронная коммерция; маркетинг; интернет-маркетинг; дропшиппинг; электронная торговля.

M.L. Kaluzhsky

**Marketing features of dropshipping in system of e-commerce**

*Annotation*: Today dropshipping wins the Internet promptly and transformed to one of the basic tools of marketing in e-commerce. Marketing features, mechanisms and value dropshipping in the conditions of network economy of the XXI century reveal in article. The author carries out the comparative analysis of institutional development dropshipping in the USA, China and Russia.

Калужский Михаил Леонидович (Омск, ОмГТУ, frsr@inbox.ru)

**Сущность дропшиппинга**. Сокращение числа традиционных посредников в электронной коммерции привело к появлению нового вида предпринимательской деятельности – дропшиппинга. Этот вид сочетается практически со всеми формами электронной коммерции, представляя собой одновременно систему коммерческих взаимоотношений и новый способ построения сетей товародвижения в маркетинге.

*Дропшиппинг* (от англ. *Dropshipping* – «*прямая поставка*») – вид коммерческой деятельности, заключающийся в продаже товаров поставщика независимыми посредниками. Такие продажи осуществляются без передачи посредникам физических товаров, прав собственности на них, а иногда и без юридического оформления отношений.

Посредник (*дропшиппер*) продает товары поставщиков от своего имени, оформляя заказ на поставку после получения оплаты от покупателей. Затем деньги переводятся поставщику, который сам отгружает товар клиенту.[1] Поскольку дропшиппинг не требует большого начального капитала, то предпринимательские риски здесь минимальны. Проблемы могут возникнуть лишь в случае несоответствия товара заявленным поставщиком параметрам.

Среди прочих преимуществ дропшиппинга следует выделить следующее:

1. Дропшипперу не требуются складские помещения. Всё, что нужно для организации продаж – персональный компьютер с выходом в сеть Интернет и наличие поставщика.

2. Все заботы по отправке товара берёт на себя поставщик. Он же предоставляет номер для отслеживания почтового отправления, организует гарантийное и постгарантийное обслуживание, осуществляет замену бракованной продукции и предоставляет необходимую информацию о товаре.

3. Дропшиппер имеет возможность сотрудничать с любым количеством поставщиков одновременно. У него нет никаких ограничений ни по ассортименту предлагаемой продукции, ни по объёму, ни по географии продаж.

4. Дропшиппер, за счёт отгрузки поставщиком товаров от его имени, имеет возможность самостоятельно создавать и продвигать узнаваемые торговые марки в Интернете.

---

[1] Сленговое наименование дропшиппера в английском языке «*desk jobber*» – посредник, не покидающий своего стола.

Кроме того, дропшиппинг, как разновидность электронной коммерции, использует некоторые элементы маркетинговой технологии прямых продаж. Используя преимущества интернет-коммуникаций, он существенно упрощает технологии управления продажами.

Во-первых, как и торговые представители, дропшипперы не являются сотрудниками компании поставщика и работают за комиссионные. Однако размер комиссионных дропшиппер назначает себе сам. Поставщик лишь отгружает товар по указанному им адресу по отпускной цене после получения оплаты.

Во-вторых, как и в прямых продажах, поставщик осуществляет информационную поддержку дропшипперов и предоставляет гарантии поставок. Однако эта поддержка ограничивается возможностью получить информационные материалы на сайте поставщика и, в отдельных случаях, работой «горячей линии» для дропшипперов. Никаких тренингов, а также предоставления пробников и печатных материалов поставщика в дропшиппинге нет.

При этом в дропшиппинге отсутствуют какие-либо обязательства торгового представителя перед поставщиком. Партнёры вступают в экономические отношения в оплаты товара и получения заказа поставщиком. Дропшиппинг доступен всем желающим, обладающим минимальными навыками интернет-торговли. Здесь царит свободный рынок, а предпринимательский успех определяется уровнем маркетинговой подготовки дропшипперов.

Следует отметить, что дропшиппинг может существовать также и без гарантий со стороны поставщиков, а также без их ведома. Любой желающий может дропшиппингом без согласования с поставщиком. Для этого достаточно воспользоваться прайс-листом поставщика и изображениями его товаров. Единственный недостаток такого подхода – наличие обратного адреса поставщика на упаковке или сопроводительных документах при получении товара покупателем.

Особенность дропшиппинга в Интернете заключается ещё и в том, что гарантом надёжности дропшиппера выступает либо он сам (наличие реквизитов, горячей линии, отзывы на форумах), либо сторонний субъект (например, торговая площадка). Вне электронной коммерции такое было бы невозможно.

**Маркетинг в дропшиппинге**. Маркетинг в дропшиппинге носит в основном пассивный характер, ограничиваясь выбором торговых площадок (форумов, социальных сетей и т.д.) и информированием потенциальных потребителей об особенностях товара. В условиях свободной конкуренции и равных маркетинговых возможностей дропшиппинга, основным фактором маркетинговой эффективности становится уровень индивидуальной компетентности, практические знания и навыки дропшиппера.

Современный дропшиппинг, как разновидность электронной коммерции, имеет те же специфические особенности:

1. Реализация мероприятий, относящихся к элементу комплекса маркетинга «*Распределение*» (*Place*), является неотъемлемой прерогативой дропшипперов. Возможности поставщика здесь крайне ограничены и сводятся к построению единой логистической схемы приёма и обработки заказов. Основная нагрузка по организации продаж ложится на плечи дропшипперов. Именно они выступают в качестве активного элемента системы товародвижения в дропшиппинге, на свой страх и риск осваивая новые целевые рынки, торговые площадки, платежные системы и т.д.

Главным ограничителем в дропшиппинге является доступ аудитории к сети. Даже самый мелкий дропшиппер при удачном стечении обстоятельств может показать выдающиеся результаты. В сфере сбытовой конкуренции здесь доминируют, прежде всего, методы доведения информации до целевой аудитории. Основная часть покупателей ищет конкретный товар, и выигрывает продавец, предложение которого окажется в нужном месте в нужное время. Отсюда, главная задача сбытовой политики в дропшиппинге состоит в доведении предложения о продаже товара до максимального количества потенциальных потребителей.

2. Элемент комплекса маркетинга «*Цена*» (*Price*) является важнейшим фактором обеспечения конкурентоспособности в дропшиппинге. Основная масса покупателей обращается к электронной торговле в надежде найти более низкую цену на товар, чем в традицион-ной торговле. Причем цена является в первую очередь маркетинговым инструментом поставщика, так как дропшипперы получают товары по единым отпускным ценам.

Задача поставщика – обеспечить такую разницу между справедливой ценой в сознании потребителей и своей отпускной ценой, чтобы предлагаемый товар стал привлекателен для потенциальных дропшипперов. Далее вступает в силу закон эластичности спроса по цене и объемы закупок (продаж) будут регулироваться лишь потребительским спросом на рынке.

Такие методы управления ценами как скидки и ценовая дискриминация потребителей в дропшиппинге применяются крайне редко. Это обусловлено близостью рынка дропшиппинговой торговли к условиям совершенной конкуренции: здесь очень много продавцов, а прибыль извлекается от объёмов продаж, а не от завышения цен на единицу товара. В дропшиппинге отсутствуют возможности для манипулирования ценами. Если у дропшиппера много конкурентов, то его наценка минимальна. Если же по какой-то причине у дропшиппера мало конкурентов, то его цена близка к отпускной цене продавца.

3. Элемент комплекса маркетинга «*Товар*» (*Product*) имеет в дропшиппинге второстепенное значение. Это обусловлено тем, что дропшипперы не обладают товаром и не могут самостоятельно определять параметры предложения. Их маркетинговые возможности ограничиваются формированием торгового ассортимента.

Дропшиппер может увеличить срок гарантии или дополнить товар комплектующими (например, картами памяти или батарейками). Однако маркетинговый эффект от этих действий быстро нивелируется сопоставимыми маркетинговыми возможностями других участников рынка. С одной стороны, все конкуренты дропшиппера имеют равный доступ к товарам поставщиков. С другой стороны, потребители через сеть Интернет обладают равными возможностями по получению информации о товаре, его свойствах и конкурентных преимуществах (в том числе на тематических форумах).

Поставщик никаких дополнительных возможностей в сфере товарной конкуренции дропшипперу также не предоставляет. Таким образом, свойства товара в дропшиппинге – это инструмент маркетинга поставщика (производителя), но никак не дропшипперов. Рынок очень быстро сведёт на нет преимущества дропшиппера, сделавшего ставку на уникальность своего торгового предложения.

4. Реализация мероприятий, относящихся к элементу комплекса маркетинга «*Продвижение*» (*Promotion*) в основном является прерогативой поставщиков (производителей) товара. Дропшипперы практически не занимаются рекламными мероприятиями. Их возможности ограничиваются маркетинговым инструментарием, предоставляемым провайдерами услуг интернет-торговли (социальными сетями, торговыми площадками и т.д.). При этом частично функции по продвижению товаров принимают на себя провайдеры, продвигая

Из арсенала PR-мероприятий дропшипперы в некоторых случаях используют т.н. «вирусный маркетинг». В социальных сетях, купонных сервисах и шоурумах эта форма привлечения клиентов является доминирующей. Однако вирусный маркетинг является атрибутом соответствующих форм электронной коммерции, а не непосредственно дропшиппинга. Поэтому ассоциировать вирусный маркетинг в дропшиппингом было бы некорректно.

Обычно функции продвижения выполняет либо поставщик (производитель), либо провайдер торговых интернет-услуг. Задача поставщика (производителя) товаров заключается в том, чтобы с минимальными затратами сформировать у интернет-аудитории привлекательный образ товара. Затем товар предлагается дропшипперам, которые принимают на себя функции по реализации товара интернет-пользователям. Провайдеры услуг интернет-торговли (социальные сети, торговые площадки и т.д.) продвигают совокупное предложение продавцов с целью привлечения покупателей.[2]

Возможности проведения маркетинговых исследований в дропшиппинге также ограничены ввиду того, что дропшиппинг редко выходит за рамки малого бизнеса. В основном эти исследования сводятся к анализу потребительского спроса (по рейтингу продавцов), ана-лизу конкуренции (количество конкурентов) и ценовых категорий (через анализ предложений), а также к поиску поставщиков высокорентабельных товаров и выявлению неохваченных ниш на потребительском рынке.

---

[2] См., напр.: Сайт торговой площадки «Молоток.Ру». – http://molotok.ru.

**Инфраструктура дропшипинговой торговли**. Торговая инфраструктура дропшипинга значительно отличается от инфраструктуры в традиционной торговле. Традиционная оптово-розничная торговля осуществляется традиционными методами (расчётно-кассовое обслуживание, договорная система взаимоотношений и т.п.). Тогда как дропшипинг больше напоминает ярмарочную торговлю с большим числом розничных продавцов и покупателей. *«Размеры заказанных по Интернету партий товаров намного меньше, но их число – намного больше…»*, – отмечают Т.Кент и О.Омар.[3]

Практически никто из дропшиперов не в состоянии самостоятельно выполнять функции традиционной торговли (хранение и сортировка товара, отгрузка продукции и т.п.). Поэтому уровень развития дропшипинга определяется уровнем развития сопутствующих сервисных логистических услуг. Можно выделить три основных вида сервисных услуг, без которых современный дропшипинг невозможен: интернет-коммуникации, платёжные операции и транспортные услуги.

Дропшипинг не является определяющим фактором существования ни одного из перечисленных сервисов. Мало того, эти сервисы не способны даже отделить дропшиперов от обычных пользователей. Однако их наличие в совокупности неизбежно ведёт к возникновению и развитию дропшипинговой торговли:

1. *Интернет-коммуникации* обеспечивают в дропшипинге 99,9% всех коммуникаций с поставщиками, партнёрами и потребителями. Современный дропшипинг без Интернета существовать не может. В отношениях с поставщиками они используются для поиска поставщиков, размещения заказов и рекламаций. Причём во всех случаях дропшипер зависит от качества и своевременности выполнения заявок поставщиком.

Поиск поставщиков осуществляется на сайтах поставщиков, либо на специализированных сайтах, обобщающих конкурентные предложения. Поэтому основная нагрузка в ходе интернет-коммуникаций ложится на поставщиков. Именно они должны так построить логистическое сопровождение сделок, чтобы дропшиперы даже не задумывались о проблемах, связанных с обработкой и выполнением заявок покупателей.

Под взаимоотношениями с партнёрами понимаются взаимоотношения с логистическим сервисами в сети Интернет, обеспечивающими стабильность и предсказуемость коммерческих операций. Дропшиперы здесь тоже практически не влияют на параметры оказываемых логистических услуг, выступая в роли обычных клиентов. При этом совершенно естественно, что наибольшее количество обращений дропшиперов будет к тем сервисам, где соотношение «цена-качество» предоставляемых услуг будет оптимальным.

Лишь в сфере взаимоотношений с покупателями дропшиперы обладают полной свободой действий, не свойственной традиционной торговле. В их распоряжении находится весь спектр возможностей, предоставляемых электронной коммерцией. Поэтому покупатели сталкиваются с дропшиперами в любой сфере интернет-коммуникаций: от социальных сетей и блогов до интернет-магазинов и торговых площадок.

2. *Платёжные операции* обеспечивают в дропшипинге важнейшую функцию – обеспечение денежных потоков. В отличие от традиционной торговли денежные потоки производятся здесь в виде частных денежных переводов. В условиях российской действительности дропшипинговые сделки производятся фактически вне правового поля и налогового контроля. Выступающие в роли частных лиц (в лучшем случае – индивидуальных предпринимателей) дропшиперы не используют традиционные для юридических лиц методы осуществления денежных переводов.

Задача дропшиперов здесь заключается в предоставлении покупателям возможности использовать максимальный ассортимент наиболее популярных услуг по осуществлению де-нежных переводов. Чем шире этот ассортимент, тем будет привлекательнее для покупателя приобретение товаров у дропшиппера. Однако влиять на параметры оказываемых услуг ни один дропшиппер не способен ввиду своей индивидуальной незначительности для провайдеров финансовых услуг.

---

[3] Кент Т., Омар О. Розничная торговля. – М.: Юнити-Дана, 2007. – С. 697.

3. Транспортные услуги завершают триаду основных факторов, обеспечивающих существование дропшиппинга. Длительные сроки поставок традиционно являются существенным препятствием для привлечения покупателей в сфере электронной коммерции.[4] Максимум что могут сделать дропшипперы – предложить покупателям на выбор услуги различных транспортных провайдеров. В дропшиппинге решение этой проблемы ложится на плечи поставщиков, которые взаимодействуют с транспортными провайдерами и организуют доставку товара. При этом именно дропшипперы несут перед покупателями ответственность за своевременность поставок, выступая перед покупателями в качестве поставщиков.

Доставка товара может осуществляться с использованием услуг ФГУП «Почта России», через промежуточные склады и почтоматы, а также скоростной почтой (EMS, DHL) или транспортной компанией на выбор покупателя. Задача как дропшипперов, так и поставщиков состоит в том, чтобы предоставить потенциальным потребителям исчерпывающую информацию о стоимости, способах и условиях поставки товара. Идеальной является такая ситуация, когда проблемы доставки воспринимаются потребителями «при прочих равных» и они не задумываются об их разрешении.

Специфика дропшиппинга заключается в том, что он одинаково комфортно развивается на любой торговой платформе. Дропшиппинг вполне обходится существующими интернет-сервисами, предназначенными для некоммерческого использования частными лицами. В этом заключается базовое конкурентное преимущество дропшиппинга перед другими видами как традиционной, так и электронной коммерции.

1. *Социальные сети*. В социальных сетях дропшиппинг является едва ли не основным методом организации продаж. К примеру, подавляющая часть коммерческих групп социальной сети «Вконтакте» представляют собой разновидность дропшиппинговой торговли.

Особенность дропшиппинга в социальных сетях заключается в том, что дропшиппер выступает не в роли продавца, а в роли представителя покупателей. На первом этапе пользователь социальной сети с большим количеством «друзей» проводит PR-кампанию среди целевой аудитории. Цель кампании – привлечь потенциальных покупателей к участию в совместной покупке напрямую от производителя по низкой цене.

Затем дропшиппер аккумулирует собранные средства, снимает свою комиссию и направляет заказ, оплату за товары и данные покупателей поставщику. Отзывы удовлетворённых покупкой клиентов используются для привлечения новых покупателей. По этой схеме действуют руководители групп, объединяющих молодых матерей, любителей брендовой одежды и т.д. Главное условие её осуществления – наличие потенциальных потребителей и недоступность для них товаров по предлагаемым ценам.

2. *Сервисы коллективных покупок*. На сервисах коллективных покупок («Groupon», «Biglion», «Darberry», «Выгода.Ру» и пр.), в качестве дропшипперов выступают организаторы сервисов. Они привлекают потенциальных клиентов возможностью приобретения товаров со скидками, коллективного участия в оптовых распродажах и т.д.

В качестве объекта маркетингового воздействия дропшипперов здесь выступают не только покупатели, но и поставщики. Поставщики, в стремлении привлечь больше покупателей, вынуждены предоставлять завышенные скидки, которые сервис коллективных покупок и «продаёт» затем своим клиентам.

Отличие от других форм дропшиппинга заключается в том, что условия диктует не поставщик, а дропшиппер в лице сервиса коллективных покупок. Его коммерческий «вес» заключается в возможности привлечения большого числа покупателей. Поставщики в усло-виях конкуренции вынуждены принимать условия такого дропшиппера, чтобы не превратиться в «ценовое пугало» для покупателей.

---

[4] Кент Т., Омар О. Розничная торговля. – С. 697.

3. *Интернет-форумы*. На интернет-форумах дропшиппинг менее распространён ввиду их неприспособленности для ведения электронной торговли. Чаще всего там обсуждаются вопросы применения дропшиппинга и размещаются ссылки на интернет-ресурсы, связанные с дропшиппинговой торговлей.

4. *Интернет-сайты*. Использование интернет-сайтов в целях дропшиппинговой коммерции очень распространено в сети. Дропшиппинговая торговля на равных конкурирует с другими видами интернет-торговли: торговыми ресурсами производителей и интернет-ритейлом. В силу своей специфики, дропшиппинговая торговля через интернет-сайты ориентирована в основном на китайских производителей.

Это обусловлено тем, что только китайские производители обеспечивают достаточную прибыль для рентабельной работы такого интернет-ресурса. Однако у ориентации на Китай есть оборотная негативная сторона: низкое качество товаров и отсутствие послепродажного сервиса отталкивает часть покупателей. Здесь почти нет постоянных клиентов. Значительная часть покупателей совершает одну покупку и больше не возвращается.

5. *Торговые площадки*. На торговых площадках дропшиппинг занимает значительную долю в общем объёме продаж. Отличительной чертой дропшипперов на российских торговых площадках является большой товарный ассортимент и поставка товаров из-за рубежа. Далеко не все проекты в области дропшиппинга удачны. Однако традиционная для России низкая плата за выставление товаров делает дропшиппинг почти безубыточным видом электронной коммерции.

Кроме того, на многих торговых интернет-площадках действуют программы защиты покупателей, значительно повышающие доверие к продавцам. Например, на торговой площадке «Молоток.Ру» покупателям, пострадавшим от мошеннических действий (обмана, непоставки, несоответствия товара описанию и т.д.) продавца выплачивается компенсация в сумме до 5000 рублей.[5] Это ставит дропшипперов в один ряд с продавцами, торгующими имеющимися в наличии товарами и позволяет им на равных конкурировать с другими видами интернет-коммерции.

6. *Псевдо-шоурумы* представляют собой гибридную форму традиционной торговли и дропшиппинга. От дропшиппинга они унаследовали методы работы с поставщиками, а от традиционной торговли – наличие резервных товарных запасов и личное общение продавца (дропшиппера) с покупателями. Благодаря большому числу клиентов псевдо-шоурумы могут позволить себе создание резервных запасов высоколиквидного товара под обеспечение потенциального спроса. Причём в отдельных случаях покупатель может приобрести товар немедленно по более высокой цене.

Однако, несмотря на перечисленные особенности, псевдо-шоурумы всё равно остаются дропшипперами, поскольку выступают в качестве посредника между зарубежными торговыми площадками (AliExpress, eBay и т.д.) и конечными покупателями. Торговые предложения псевдо-шоурумов определяются ценовыми дисбалансами в рамках спроса и предложения, а также неосведомлённостью покупателей. Пока указанные факторы будут существовать, будут существовать и псевдо-шоурумы.

Лишь электронные доски объявлений не подходят для дропшиппинга в силу локальности предложения. Их особенность заключается в том, что сделка заключается в ходе личного контакта после отклика покупателя на объявление продавца. Такая форма электронной коммерции не предусматривает длительного ожидания товара после предоплаты.

В целом дропшиппинг может существовать в самых разнообразных формах электронной коммерции. И не важно, выступает дропшипер в качестве посредника между сайтами производителей и покупателями, между интернет-магазинами (самостоятельными и на тор-говых площадках) и покупателями, между покупателями и поставщиками (купонные сервисы). Значение имеет лишь то, что дропшиппинг обусловлен объективной востребованностью такого рода услуг на рынке.

---

[5] См.: Программа защиты покупателей / Открытая торговая площадка «Молоток.Ру». – http://molotok.ru/country_pages/168/0/education/pok/index.php?page=0

**Американская практика дропшиппинга.** Специфика применения дропшиппинга в США обусловлена тем, что дропшиппинг как форма маркетинга сформировался там ещё в начале XX века. Одно из первых упоминаний о дропшиппинге относится к 1927 году, когда американские маркетологи Г.Мейнард, В.Вейдлер и Т.Бекман дали его подробное описание.[6]

Согласно их трактовке: «*Дропшиппер ... имеет офис, но не склад, так как он не владеет физическими товарами ... он принимает на себя право собственности на товар и ответственность за отгрузку*».[7] При этом отмечалось, что «*дропшиппер является самым важным оптовым торговцем ограниченной функции обслуживания*».[8]

По данным четвертого издания книги Г.Мейнарда, В.Вейдлера и Т.Бекмана «Принципы маркетинга» (1946) в 1939 году на рынке США активно действовало свыше 1000 дропшипперов с продажами 475 млн. долларов, что составляло около 2% общего объёма оптовой торговли. Основная часть (около 90%) дропшипперов торговала углём и коксом. Среднестатистическая доля трансакционных издержек дропшипперов, обусловленная отказом от дорогостоящих функций (складирования, погрузочно-разгрузочных работ и т.д.), составляла около 6,4% общего объёма продаж против 21,0% у полнофункциональных оптовых торговцев.[9]

Естественным ограничением дропшиппинга выступали минимальные партии отгружаемого товара и оригинальная упаковка товаров с указанием реквизитов поставщика. Поскольку дропшипперы не обладали складскими помещениями и соответствующим персоналом, то они были вынуждены совершать сделки только на такие партии товара, которые соответствовали требованиям поставщика (например, вагон угля). С другой стороны дропшипперы не могли торговать товарами, на которых были указаны реквизиты поставщика во избежание прямых контактов с ним покупателей.

Сдерживало развитие дропшиппинга и негативное отношение к дропшиппингу многих поставщиков того времени, которые неохотно шли на сделки с дропшипперами. Их не устраивало стремление дропшипперов перекладывать важнейшие функции оптовой торговли на поставщиков. В первую очередь это относилось к дополнительному хранению зарезервированного товара и риску, связанному с устареванием товара (например, сельскохозяйственной продукции). Дропшипперы, с целью противодействия неприятию поставщиками, вынуждены были закупать товар заранее, откладывая поставку до его перепродажи.

Однако, несмотря на высококонкурентные условия существования, дропшиппинг в послевоенный период значительно увеличил свою долю в общем объёме отраслевой оптовой торговли. Так к 1955 году в США насчитывалось уже около 2600 дропшипперов с общим объёмом продаж $2,2 млрд. долларов.

Следует отметить стабильную тенденцию сближения показателей трансакционных издержек у дропшипперов и полнофункциональных оптовых торговцев. Если в 1939 году трансакционные издержки составляли около одной трети соответствующих издержек полнофункциональных оптовых торговцев, то к 1954 году данный показатель превысил половину. Это объясняется развитием коммуникативных каналов и доступностью дропшипинговых технологий для всех участников рынка.

Позже произошла конвергенция дропшиппинга и полнофункциональной оптовой торговли. Американские маркетологи Т.Бекман, Н.Ингл и Р.Баззелл в 1959 году отмечали: «*Обзор 126 оптовых торговцев показал, что приблизительно 23% их продаж были произведены*

---

[6] Maynard H.H., Weidler W.C., Beckman T.N. Principles of marketing. – New York: Ronald Press, 1927.
[7] Цит. по: Scheel N.T. Drop Shipping as a Marketing Function: A Handbook of Methods and Policies. – Westport (CT): Praeger Publishers, 1990. – P. 4.
[8] Там же.
[9] Там же. – P. 5.

*с помощью дропшиппинга*».[10] Их исследования показали абсолютное доминирование дропшиппинга в промышленных поставках сырья и оборудования в США к началу 1960 годов.

Дропшиппинг после 1960-х гг. постепенно превращается из самостоятельной разновидности оптовой торговли в её инструмент, применяемый всеми участниками рынка исходя из экономической целесообразности. Не случайно видный американский исследователь дропшиппинга Николас Шиль назвал свою монографию «Дропшиппинг как маркетинговая функция».[11] Это не было похоже на современный дропшиппинг, поскольку речь тут шла об оптовой торговле в сфере «B2B», а не о розничной торговле потребительскими товарами в сфере «B2C». Такая торговля могла существовать только в условиях недостаточно развитых информационных технологий, когда потенциальные потребители не имели доступа к информации о поставщиках.

На начальном этапе дропшипперы получали прибыль за счёт монопольного владения информацией о поставщиках, товарах и условиях поставок. В то же время они были весьма конкурентоспособны в сравнении с полнофункциональными оптовыми торговцами из-за низких трансакционных издержек при поставке товара. Положительная роль дропшипперов в условиях конкурентной экономики заключалась в том, что они естественным образом ограничивали наценки в оптовой торговле.

Однако последующее развитие информационных коммуникаций и биржевой торговли привело к растворению дропшиппинга в традиционной оптовой торговле. Дропшипперы утратили своё конкурентное преимущество, основанное на лучшем знании рынка и более высокой мобильности в сравнении с другими участниками рынка. Их источники информации стали общедоступными, а методы работы вошли в арсенал конкурентов. Дропшиппинг как самостоятельный вид коммерческой деятельности утратил свою актуальность и оказался забыт на многие годы.

Ситуация кардинально изменилась с появлением Интернета и электронной коммерции, которые не просто снизили трансакционные издержки и сделали интернет-торговлю доступной для миллионов энтузиастов. Они предоставили интернет-пользователям то же конкурентное преимущество, которым обладали оптовые дропшипперы на заре становления дропшиппинга – коммуникационную мобильность.

Специфика возрождённого на новой основе американского дропшиппинга (а также ориентирующегося на него дропшиппинга в Европе, Японии, Канаде и Австралии) заключается в соблюдении институциональных традиций, правил и обычаев организации продаж. Современный американский подход ориентирован на традиционные торгово-закупочные отношения, рассматривающие дропшиппинг как функцию маркетинга, а не как самостоятельный вид электронной коммерции.

Отчасти это можно объяснить особенностями американской экономики. Здесь нет экономической базы в виде повышенной доли прибыли для существования дропшиппинга в качестве самостоятельного вида экономической деятельности. Разница между отпускной ценой поставщиков и розничной ценой на рынке невелика. Кроме того, поставщики привязаны к сложившейся экономической инфраструктуре в виде ритейлеров и оптово-розничных посредников.

Сегодня на рынке США можно встретить самые различные виды дропшиппинга. Однако доминирующим направлением в американском дропшиппинге является не продажа товаров конечным потребителям, а иерархическое развитие инфраструктуры дропшиппинга. Такое развитие заключается в формировании на рынке крупных дропшипперов, оказывающих соответствующие услуги мелким дропшипперам.

Мелкие дропшипперы в силу своей неорганизованности не в состоянии самостоятельно осуществлять поиск поставщиков и предлагать им привлекательные условия сотрудничества. Этим пользуются прикрывающиеся дропшиппингом сервисные структуры, которые оказывают посреднические услуги, как поставщикам, так и мелким дропшипперам, фактически монополизируя сферу межфирменных коммуникаций.

---

[10] Там же. – P. 6.
[11] Scheel N.T. Drop Shipping as a Marketing Function: A Handbook of Methods and Policies. (1990).

В качестве примера можно привести одну из крупнейших в США дропшиппинговую интернет-компанию «Doba», основанную в 2002 году.[12] Эта компания специализируется на комплектовании ассортимента интернет-магазинов по схеме дропшиппинга. Каталог предлагаемых к продаже товаров компании «Doba» по состоянию на 02.09.2012 г. насчитывал 1.447.353 товара в 1.500 категориях (около 8.000 брендов) от 165 поставщиков.

Доступ к каталогу предоставляется дропшипперам на платной основе. Ежегодная абонементная плата составляет от 599,50 долларов до 899,50 долларов в зависимости от пакета предоставляемых услуг. Все сделки дропшипперов с поставщиками производятся через интерфейс сайта компании. Через веб-интерфейс дропшипперы получают также готовые описания и фотографии товара.

Функции дропшипперов, сотрудничающих с компанией «Doba» сводятся к размещению торговых предложений в сети Интернет, а также в сборе и обработки заказов. При этом дропшипперам категорически запрещается самостоятельно контактировать с поставщиками. Несмотря на то, что по данным журнала «Inc.»[13] компания «Doba» входила в 2007-2009 гг. в число наиболее быстрорастущих компаний США, источники такого роста вызывают сомнения по следующим причинам:

Во-первых, прибыль компании «Doba» формируется за счёт абонементной платы дропшипперов. Система отношений с дропшипперами построена так, что абонементная плата списывается авансом с банковского счёта в безакцептном порядке и не зависит от показателей продаж.

Во-вторых, огромный ассортимент товаров имеет обратную сторону – объёмы продаж отдельных товаров крайне невелики. Кроме того, поставка почти 1,5 млн. товаров 165 поставщиками свидетельствует о том, что поставщики не являются конечными производителями. Мало того, их отпускные цены достаточно высоки и не могут конкурировать с ценами сезонных распродаж.

В-третьих, особенность экономики США заключается в том, что основное производство потребительских товаров расположено за пределами США (в Мексике, Китае, Таиланде, Египте и т.д.). Соответственно ценовое предложение внутри США не способно конкурировать с предложениями от производителей из этих стран.

Можно предположить, что компания «Doba» не обладает постоянной клиентской базой, выступая в качестве «ловушки для новичков». Вместе с тем, именно такие сервисы «посредников для посредников» показывают в США наилучшие финансовые результаты.

Описанная тенденция соответствует общему направлению развития постиндустриальной сервисной экономики в развитых странах. Не случайно некоторые авторы считают оказание информационных услуг в сфере электронной коммерции (и дропшиппинга в частности) самостоятельным видом продаж через Интернет. Речь идёт о выделении новой разновидности электронных розничных продаж, под которой понимается *«продажа или распространение информации о том, где можно продать или купить определённый товар»*.[14] Однако в условиях падения совокупного спроса и экспансии на мировых рынках китайских производителей описанный подход не выглядит достаточно конкурентоспособным.

**Азиатская практика дропшиппинга**. Азиатская практика дропшиппинга заслуженно ассоциируется с Китаем. У китайского подхода к организации дропшиппинга есть свои причины и отличительные особенности. Несмотря на внешнюю схожесть в методах и инструментарии, китайский дропшиппинг кардинально отличается от американского.

Именно эти отличия определяют во многом успехи и экспансионистские стратегии китайских товаропроизводителей. Объясняется данное обстоятельство тем, что в Китае

---

[12] Сайт дропшипинговой компании «Doba» (США). – http://www.doba.com.
[13] Сайт интернет-журнала «Inc.» – http://www.inc.com/about/index.html
[14] Кент Т., Омар О. Розничная торговля. – С. 695.

дропшиппинг – доминирующая форма продаж в условиях экспортно-ориентированной экономики. Ему не приходится бороться с устоявшимися традиционными видами торговли.

Так, в китайской модели дропшиппинга реклама не играет определяющей роли в организации продаж. Её место занимает цена товара, которая является главным стимулом для привлечения как дропшипперов, так и конечных потребителей. Именно цена делает китайские товары привлекательными для покупателей по всему миру, несмотря на сомнительное качество товаров и отсутствие послепродажного обслуживания.

В Китае дропшиппинг не сразу сформировался в качестве самостоятельной разновидности торговли, а стал логическим продолжением общего процесса становления оптово-розничной торговли. Можно выделить несколько стадий этого процесса, обусловленных особенностями китайской экономики.

1 стадия. *Челночная торговля*. Пик развития челночной торговли (начало 1990-х гг.) был сопряжён с выводом крупнейшими мировыми производителями производственных мощностей в Китай. В стране появились новые, не совсем легальные с точки зрения соблюдения прав интеллектуальной собственности, производственные возможности. Низкая стоимость рабочей силы обусловила высокую конкурентоспособность продукции.

При этом контрафактная продукция не могла легально поставляться на внешние рынки. Контролируемая крупными зарубежными производителями торговая инфраструктура не хотела работать с такими товарами. Проблема решалась через китайские рынки по всему миру, а в России и СНГ ещё и с привлечением отечественных «челноков».

Если в начале 1990-х гг. оптово-розничная торговля в Китае начиналась с рыночных палаток, то к концу 1990-х гг. китайские поставщики использовали развитую сеть шоурумов, накопительных складов и комплексы транспортной логистики. Однако во второй половине 1990-х гг. челночная торговля постепенно начала изживать себя за счёт проникновения китайских товаров на мировые сети через традиционную торговую инфраструктуру и негативного отношения покупателей к некачественной контрафактной продукции.

У челночной торговли были недостатки, присущие магазинной торговле. К ним можно отнести большую длительность торгового цикла, непредсказуемость спроса и низкую квалификацию продавцов. Поэтому челночная торговля могла конкурировать с традиционной торговлей только в низшем ценовом сегменте. Однако она сформировала институциональную основу для дальнейшего перехода к дропшиппингу.

2 стадия. *Продающие интернет-сайты*. Эта стадия началась во второй половине 1990-х гг. и завершилась в начале 2000-х гг. Китайские посредники открывали независимые интернет-магазины и рассылали сообщения в виде несанкционированных рассылок (спама). Отдача от таких рассылок была невелика и не превышала десятых долей процента. Покупателей отталкивала малоизвестность продавцов и отсутствие каких-либо гарантий поставок. Однако попытки не прекращались ввиду практически бесплатной возможности рассылки информации по миллионам e-mail адресов со всего мира.

Вторая стадия продлилась недолго. Её заслугой стало появление в Китае интернет-ориентированных продавцов, не понаслышке знакомых с особенностями электронной коммерции. Кроме того, большую роль сыграло присутствие китайских продавцов на торговой площадке «eBay» и открытие дочерних торговых площадок «eBay» в Китае (www.ebay.cn) и Гонконге (www.ebay.com.hk). Полученный ими опыт впоследствии был использован для развития китайский торговых интернет-площадок.

3 стадия. *Дропшиппинговые компании и торговые площадки*. Переход к третьей стадии связан с появлением в Китае интернет-инфраструктуры, обеспечивающей техническую поддержку дропшиппинговой торговли. Электронная коммерция в Китае не только скопировала созданные в США технологии дропшиппинговой торговли, но и наполнила их новым содержанием. Основное отличие китайской модели дропшиппинга состоит в её ориентации на нужды и потребности потенциальных дропшипперов. В этом же кроется причина успеха китайской модели, превратившей дропшиппинг в один из важнейших видов продвижения товаров в электронной коммерции.

Во-первых, оплата на китайских дропшипинговых сайтах никогда не взимается с потребителей. Дропшипперы, зарегистрированные в качестве продавцов платят на торговой площадке «Alibaba» довольно большой взнос – 2.999 долларов США в год.[15] Описанный взнос предназначается для продавцов, а не для покупателей. Большая посещаемость китайских интернет-площадок ведёт к тому, что подавляющее число торговцев там являются дропшипперами, несмотря на чрезмерно высокие сборы.

Китайские системы распределения продукции среди местных дропшипперов основываются на сети представительств производителей, работающих по принципу «склад-магазин». В таких представительствах дропшипперы получают свежие прайс-листы, приобретают товары за наличный расчёт и туда же обращаются за заменой бракованных товаров.[16]

Дропшипперы могут продавать товары любым способом и в любом месте. Часть из них предпочитает торговать на китайских торговых площадках или на международных аукционах. Некоторые создают специализированные сайты в Интернете. Отличительная черта такого подхода заключается в полном отсутствии входного барьера для новых участников. Любой желающий способен стать дропшиппером, абсолютно бесплатно получив всю необходимую информацию и доступ к прямым поставкам.

В Китае, где функции производства и сбыта зачастую разделены между контрагентами, сложилась система сбыта, в рамках которой инвесторы выкупают большие партии высоколиквидных товаров у производителей, а затем перепродают их дропшипперам и конечным потребителям от своего имени. Иначе говоря, если ценовой дисбаланс между отпускной и розничной ценой достаточно велик, то сбытовая инфраструктура выстаивается сама собой. Это одно из важных преимуществ глобального рынка и интернет-торговли.

4 стадия. *Интернационализация китайской модели дропшиппинга*. Успешное развитие китайской модели дропшиппинга не могло не повлиять на дропшиппинг в других странах. Неизбежным следствием проникновения китайских товаров на зарубежные рынки стало распространение китайских технологий продаж.

Большим стимулом для этого является мощнейшая логистическая поддержка, оказываемая экспорту китайских товаров. Например, логистическая служба «PFC express» компании «Shenzhen Royal International Logistics Co., LTD» обеспечивает скидки поставщикам при почтовой отправке товаров в сумме от 60% до 80% от стандартных почтовых тарифов.[17]

В результате меняется сама схема поставок. На смену китайским дропшипперам приходят специализированные сервисы по приёму и обработке заказов через Интернет. Экспансия китайской модели дропшиппинга распространяется далеко за пределы Китая. Теперь в качестве дропшипперов китайской продукции всё чаще выступают зарубежные продавцы, предлагающие товары от своего имени. Они лучше знают специфику своих рынков и вызывают больше доверия у потенциальных покупателей.

В качестве примера такого сервиса можно привести сайт «DGS» гонконгской компании «Dropshipping Global Services Limited», специализирующейся на торговле брендовой электроникой («Sony», «Panasonic», «Samsung», «HTC» и др.).[18] Эта компания оказывает полный комплекс услуг по обработке, доставке и отслеживанию поставок для зарегистрированных дропшипперов, выступающих в качестве дилеров компании.

Одновременно китайские дропшипперы начинают осваивать новую для себя сферу «B2B», оказывая крупным зарубежным заказчикам посреднические услуги по размещению заказов на китайских предприятиях и прямым закупкам товаров. Получается своеобразный «дропшиппинг наоборот», когда заказчиком (субъектом) дропшипинговых услуг выступает покупатель, а объектом – товары китайских производителей. Тогда как китайские дропшипперы всё больше ориентируются на местный потребительский спрос.

---

[15] Электронная торговая площадка «Alibaba». – http://www.aliexpress.com.
[16] На сленге китайских дропшипперов эти представительства называются «маркетинг».
[17] Сайт логистической службы «PFC Express». – http://www.parcelfromchina.com.
[18] Сайт компании «Dropshipping Global Services Limited». – https://www.dropshipgs.com.

Китайская модель дропшиппинга в России представлена сегодня не только отечественными продавцами китайских товаров. Крупнейший в Китае сайт «AliExpress» с апреля 2010 г. доступен в России на русском языке.[19] Эта торговая площадка не только предоставляет возможность прямых заказов товаров по оптовым ценам китайских производителей. Интерфейс сайта позволяет осуществлять немедленную оплату покупок из России с помощью банковских карт «VISA» и через российскую платёжную систему «Qiwi».[20] Информационную поддержку сайта, а также ответы на вопросы покупателей осуществляют около десятка групп российской социальной сети «Вконтакте».[21]

Однако распространение китайского варианта электронной коммерции не препятствует её развитию в других странах и в России в частности. Важным следствием интернационализации дропшиппинга в Китае является то, что китайские производители утратили контроль над инфраструктурой дропшиппинга. В результате развития дропшиппинговой торговли не только китайские посредники, но и российские предприниматели могут заключать прямые сделки с китайскими производителями.

**Состояние и перспективы дропшиппинга в России.** Из-за периферийности отечественной торговой инфраструктуры дропшиппинг пришёл в Россию с запозданием минимум на несколько лет. Отечественные дропшипперы, выходя на зарубежных поставщиков, столкнулись с уже сформировавшимися институтами, методами и формами дропшиппинга. С одной стороны, это предопределило специфику и направления развития отечественного дропшиппинга. С другой стороны, это упростило процесс становления дропшиппинга в России.

С началом формирования технологической базы дропшиппинга в России практически сразу начался процесс его бурного становления. Можно выделить сразу несколько взаимосвязанных причин, обусловивших быстрое распространение дропшиппинга:

1. *Дисбаланс между розничными ценами* в России и оптово-розничными ценами в других странах (Китае, Европе и США). Благодаря почтовой доставке и интернет-коммуникациям появилась возможность заказывать товары напрямую за рубежом. Это не могло не сказаться на снижении трансакционных издержек и ускорении товарооборота. Коммерческая эффективность сделок стала гораздо выше, чем в традиционной торговле.

Сначала интернет-пользователи заказывали товары для себя, а затем, используя приобретённый опыт, предлагали эти же товары по схеме дропшиппинга через Интернет. Опыт, приобретённый в процессе интернет-покупок на интернет-аукционах, очень быстро конвертировался в опыт интернет-продаж.

2. *Развитие почтовой логистики* значительно облегчило отслеживание почтовых отправлений. С середины 2000-х гг. практически все почтовые администрации стран мира предоставляют возможность отслеживания прохождения почтовых отправлений по идентификационному номеру через Интернет.[22] Это значительно упростило взаимодействие дропшипперов с покупателями в условиях хронического несоблюдения сроков доставки почтовых отправлений. Достаточно сообщить покупателю идентификационный номер отправления, и он сам может наблюдать за его прохождением или получить необходимую информацию по бесплатной горячей линии.

3. *Развитие платёжных инструментов* решило проблему обеспечения надёжности финансовых расчётов между поставщиками, дропшипперами и продавцами. В 1996 г. в России появилась первая платёжная система «Anelik», снизившая стоимость денежных переводов с 12% у платёжных систем «Western Union» и «Money Gram» до 1,5-3% для сумм от 100 долларов США и ниже.[23] Немного позже к ней добавились «Contact» (1999) и «Unistrim» (2001). Это открыло дорогу к рынку дропшиппинга частным предпринимателям с небольшим оборотом.

---

[19] Русская версия сайта «AliExpress». – http://ru.aliexpress.com.
[20] Сайт платёжной системы «Qiwi». – http://qiwi.ru.
[21] См., напр.: Официальная группа Aliexpress.com . – http://vk.com/official_aliexpress.
[22] См. напр.: Сайт ФГУП «Почта России». – http://www.russianpost.ru.
[23] См., напр.: Сайт платежной системы «Anelik». – http://www.anelik.ru.

Однако подлинным прорывом в интернетизации международных платежей стало появлением на российском рынке международных платёжных сервисов «Moneybookers» (Великобритания), «PayPal» (США), которые позволили покупателям и продавцам совершать платежи, не выходя из дома. Другим их преимуществом является наличие «Программ защиты покупателей», делающих покупки в Интернете абсолютно безопасными для покупателей.

4. *Развитие электронных коммуникаций* обеспечило дропшипперов широчайшим ассортиментом приёмов и методов продвижения товаров в сети Интернет. Благодаря минимизации затрат на продвижение они получили возможность широкого охвата огромных целевых аудиторий потенциальных потребителей. Это привело к тому, что дропшиппинг превратился в один из важнейших видов электронной коммерции в социальных сетях и на электронных торговых площадках. Важным следствием развития электронной коммерции стало повышение доступности дропшиппинга вплоть до практически полной отмены «входных барьеров» для новых участников рынка.

В России развитие дропшиппинга начиналось с социальных сетей, интернет-аукционов и интернет-сайтов. Все три направления развивались параллельно. По существу, здесь речь можно вести о трёх отдельных разновидностях каналах сбыта, направленных на три отдельные целевые аудитории. У каждого из этих каналов есть своя логика внутреннего развития, свой потенциал и перспективы роста.

1. Социальные сети стали инструментом привлечения клиентов в сообщества, заинтересованные в коллективном приобретении товаров напрямую от зарубежных поставщиков. Крайней формой такой самоорганизации стали впоследствии псевдо-шоурумы.

2. Интернет-аукционы позволили дропшипперам получить готовую целевую аудиторию для продвижения товаров. Эволюционирование интернет-аукционов в электронные торговые площадке привело к эволюционированию наиболее успешные дропшипперов во владельцев специализированных интернет-магазинов на торговых площадках.

3. Дропшипинговые интернет-сайты, получившие на первоначальном этапе широкое распространение в Рунете, постепенно утрачивают доминирующую роль в электронной коммерции. Они не могут конкурировать с электронными торговыми площадками ни по размеру целевой аудитории, ни по степени доверия потребителей, ни по уровню сервисных услуг.

В целом российская модель дропшиппинга стала следствием институционального развития отечественной электронной коммерции и электронных коммуникаций. Её нельзя назвать уникальной или обладающей какими-то специфическими национальными особенностями. Для формирования собственной модели дропшиппинга в России не сложились экономические условия: здесь нет такого платёжеспособного потребительского спроса как в США или такого товарного предложения как в Китае.

Однако у дропшиппинга в России есть и свои существенные преимущества. Виртуальность интернет-пространства отменяет многие ограничения традиционной торговли. В результате абсолютно все дропшипперы обладают равными возможностями как для освоения внутрироссийского рынка, так и зарубежных рынков в рамках сложившихся глобальных институциональных отношений.